# Estimation of Cooper pair density and its relation to the critical current density in hole doped high-$T_c$ cuprate superconductors

Nazir Ahmad, S. H. Naqib*

Department of Physics, University of Rajshahi, Rajshahi 6205

*Corresponding author; Email: salehnaqib@yahoo.com

**Abstract**

Hole concentration in the $CuO_2$ plane largely controls all the electronic properties in the normal and superconducting states of high-$T_c$ cuprates. The critical current density, $J_c$, is no exception. Previous hole content dependent studies have demonstrated the role of intrinsic depairing current density in determining the observed critical current density in copper oxide superconductors. It is also widely agreed upon that the temperature and magnetic field dependent vortex pinning energy plays a major role in determining the $J_c$ of a system. This pinning energy depends directly on the superconducting condensation energy. Superconducting condensation energy, on the other hand, is proportional to the Cooper pair density (superpair density), which is found to be highly dependent on the hole concentration, $p$, within the $CuO_2$ plane. We have calculated the Cooper pair density, $\rho_s$, of YBCO (Y123), a typical hole doped cuprate, as a function of $p$, in this study. A triangular pseudogap (PG), pinned at the Fermi level, in the quasiparticle spectral density has been considered. The low-temperature critical current density of a number of Y(Ca)BCO superconductors over wide range of compositions and hole concentrations have been explored. The normalized values of the superpair density and the critical current density exhibit a clear correspondence as the in-plane hole content is varied. This systematic behavior provides us with strong evidence that the critical current density of hole doped cuprates is primarily dependent on the superpair density, which in turn depends on the magnitude of the PG energy. The agreement between the estimated $p$-dependent superpair density and the previously experimentally determined superfluid density of Y(Ca)BCO is quite remarkable.

## 1. Introduction

The phenomenon of superconductivity is now a considerable focus of attention to the scientific community as one of the most promising technologies that can address the global energy crisis by facilitating energy utilization with minimal loss and high degree of efficiency. The possibility of every large scale practical application of superconductivity depends primarily on the maximum current density which superconductors can carry (the critical current density, $J_c$), in some way or other. The value of losses incurred within the superconductors, the maximum magnetic field strength in which superconductors can be used, etc are the other important factors all intimately linked with $J_c$. All these factors are directly related to the pinning ability of the quantized magnetic flux lines (magnetic vortices) of superconductors in the superconducting state. Maximizing $J_c$ and the magnetic field under which the superconductor can perform, as well as minimizing losses, have been important goals of scientists working in the field of applied superconductivity.

High-$T_c$ cuprates discovered more than three decades back [1], belongs to the group of the most promising superconductors capable of sustaining high critical currents under high applied magnetic fields [2] even though serious theoretical and technical challenges remain [2, 3]. The initial enthusiasm surrounding the vision for extremely high critical current density, powerful magnets, motors, generators, and loss-less transmission lines working at liquid nitrogen temperatures (77 K) was based on the high superconducting transition temperatures of cuprates. The belief that a high-$T_c$ itself ensures a high $J_c$ is oversimplified. In reality applications at 77 K have turned out to be much more challenging than at 4.2 K, irrespective of the values of $T_c$ and the upper critical field, $H_{c2}$. Strong electronic correlations, $d$-wave order parameter, quasi-two dimensional structural features leading to high level of structural and electronic anisotropy, and small correlation length in high-$T_c$ cuprates make these systems fascinating materials with diverse and competing electronic orders [4 – 7]. These complexities, at the same time, hinders any attempt to make such materials useful, which involves compromises among conflicting requirements, defining the parameters of merit depending on the operating conditions and also on the specific form of the application [2, 3]. Since the early days of superconductivity in cuprates, it was realized that the effect of strong thermal fluctuations of vortices will make pinning a challenging problem at and above the boiling point of liquid nitrogen. The supercurrent carrying capability becomes limited by fluctuations of the order parameter and thermally activated hopping of the flux lines. The optimized condition is expected to be found in samples with high superfluid density and low anisotropy factor [2, 3, 8]. It also appeared that the low-$T$ $J_c$ might be mainly governed by the intrinsic superconducting parameters and the performance cannot be improved significantly via extrinsic modifications like introducing defects as pinning centers, especially in case of optimally doped compounds with maximum possible $T_c$ [2, 9, 10]. In a previous study we have gathered indications that indeed the intrinsic depairing current density sets the value of the experimental $J_c$ to a large degree [9].

Due to its comparatively high superfluid density and the lowest level of structural and electronic anisotropy, YBa$_2$Cu$_3$O$_{7-\delta}$ (Y123 or YBCO) remains among one of the most promising high-$T_c$ cuprates for potential applications. In this study we will focus on the zero-field critical current density at zero temperature, as a function of hole content for YBCO and Y$_{1-x}$Ca$_x$Ba$_2$Cu$_3$O$_{7-\delta}$ (Y(Ca)BCO) over a wide range of compositions. Zero-field and zero-temperature critical current density is expected to be dominated by the intrinsic effects as maximally developed superconducting (SC) energy gap and order parameter mask extrinsic effects to a large extent. Ca substituted compounds have been used because fully oxygenated ($\delta \sim 0$) YBCO is slightly overdoped (OD) and full oxygen loading is difficult; deeply OD regions can be accessed when trivalent Y$^{3+}$ is replaced by divalent Ca$^{2+}$ [9, 11 – 13].

It is known that the supercurrent circulates due to *phase angle twist* (measured through $\nabla\theta$) of the SC order parameter. This twist, in turn, depends on the *phase stiffness* of the SC compound. The phase stiffness, on the other hand, varies linearly with the superfluid density [14]. Therefore, from intrinsic consideration, the $J_c$ of a superconductor is expected scale with the superfluid density or the superpair density for that matter.

A number of prior studies revealed that a pseudogap (PG) correlation, competing with superconductivity, depletes superfluid density most effectively [7, 15, 16]. This arises because of the removal of the low-energy quasiparticle (QP) spectral weight around the Fermi level due to the formation of the PG. In this paper we intend to investigate this proposal via the calculations of superpair density for Y(Ca)BCO within a triangular PG scenario pinned at the Fermi level [17, 18]. The superpair density as function of hole content, $\rho_s(p)$, has been calculated from the previously estimated [10, 18 – 20] $p$-dependent PG energy scale, $E_g(p)$, using the triangular PG model. The $p$-dependent zero-field and zero-temperature critical current density, $J_0(p)$, shows clear correspondence to $\rho_s(p)$. These are the central results of this investigation.

Rest of the paper is organized as follows. Section 2 comprises of the description of the methodology used to calculate the superpair density within a simple triangular PG model. Section 3 deals with the correspondence between $\rho_s(p)$ and $J_0(p)$. Results are discussed and finally, conclusions are drawn in Section 4.

## 2. Theoretical methodology for calculation of $\rho_s(p)$

The PG is a suppression of electronic density of states (EDOS) near the Fermi level. One of the distinct features of this gap in the QP energy spectrum is that it has states non-conserving character [21 – 23] and does not show any distinct coherence peak-like feature as observed at the onset of phase coherent superconductivity. Variety of earlier studies have demonstrated the success of a states non-conserving triangular PG model pinned at the Fermi energy to explain diverse class of normal and SC state experimental results including temperature dependent resistivity [24, 25], bulk magnetic susceptibility [18, 26], impurity induced magnetic behavior [17, 27, 28], electronic heat capacity [21], and NMR Kinight shift data [29]. The schematic diagram of such a simple triangular PG is shown in Fig. 1 below.

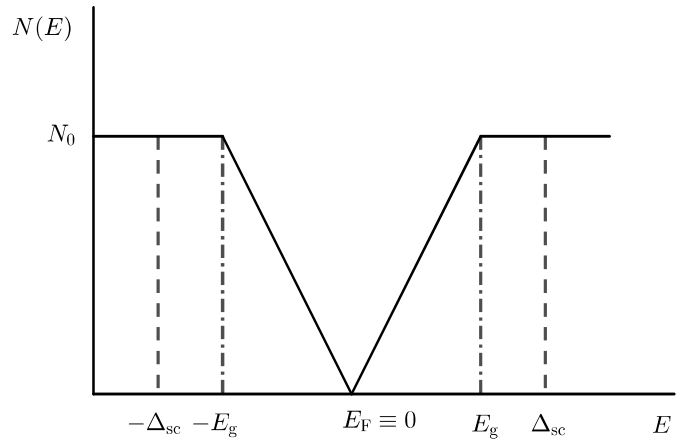

Figure 1: The linearly vanishing triangular pseudogap with energy $E_g$. $\Delta_{sc}$ marks the amplitude of the zero-temperature superconducting gap.

Figure 1 also shows the magnitude of the superconducting gap (SCG). It should be noted that the SC coherence peaks at either side of the Fermi level have not been shown in this figure. These coherence peaks are formed in the EDOS at the expense of condensed (coherently paired) charge carriers residing within the low-energy electronic density of states. The PG depletes these low-energy EDOS and reduces the superconducting condensate. In other words, in the absence of the PG, the QP spectral weight under the coherence peaks would have been much larger since in this case there would be a lot more QP spectral weight available at low-energies to take part in the SC pairing condensate.

A clear understanding of the nature of the spectral gaps remain as one of the most challenging problems in hole doped cuprates. Contrary to conventional Fermi-liquid superconductors where the gap in the QP spectral density around the Fermi level vanishes at the SC critical temperature, as described by the Bardeen-Cooper-Schrieffer (BCS) theory [14, 30], in cuprates an energy gap exists much above the critical temperature $T_c$ in the underdoped (UD) and optimally doped (OPD) compounds [11, 18, 19, 22, 31 – 34]. Distinguishing this normal state PG near $T_c$ from the superconducting (coherence) energy gap is a challenging issue. There is growing evidence that the PG is distinct from the SCG and the characteristic PG temperature,

$T^*(p)$ goes below the $T_c(p)$ dome in the slightly OD side of the phase diagram and terminating ($T^*(p) = 0$ K) at a critical hole concentration, $p_c \sim 0.19$ [11, 18, 19, 22, 31 – 35].

It has been observed from a variety of experimental probes [36, 37] that the magnitude of the SCG does not vary much over an extended region of the hole content, particularly in the region from slightly OD to moderately underdoped parts of the T-p electronic phase diagram. In this particular region, the magnitude of the SCG remains larger than the PG and the spectral gaps conform to the schematic diagram shown in Fig. 1. For simplicity, we will perform our calculations of superpair density mostly in this region. Within the simple d-wave SCG formalism, $2\Delta_{sc}/k_B T_c = 4.28$ [38]. In the strong coupling regime (prevalent to the deeply UD compounds), this ratio tends to increase. This is because $T_c$ goes down in the underdoped region, but $\Delta_{sc}$ stays high [36, 37]. In fact, over the range of hole content considered here, $\Delta_{sc} = 2.14 k_B T_{c0}$, to a reasonable approximation [36, 37] for YBCO. $T_{c0}$ denotes the maximum SC transition temperature at the optimum doping ($p_{opt} = 0.16$). For YBCO, $T_{c0} = 93$ K. We discuss the possible implications of this approximation in Section 4.

To be specific, the EDOS profile centered at the Fermi energy as shown in Fig. 1, can be modeled as,

$$N(E) = N_0(E/E_g) \quad \text{for } E \leq E_g$$

(1)

$$= N_0 \quad \text{for } E > E_g$$

where $N_0$ is the EDOS in the flat region outside the spectral gaps.

By definition, the superpair density can be expressed as,

$$\rho_s(p) = \langle N(E_F) \rangle \Delta_{sc} \qquad (2)$$

where $\langle N(E_F) \rangle$ is the normal state average electronic density of states pinned at the Fermi-level. It should be noted that, for the conventional superconductors with weakly energy dependent EDOS around the Fermi energy, this quantity is almost identical to the $N(E_F)$. Due to the symmetrical nature of the EDOS above and below $E_F$ within a window of energy of width $\pm \Delta_{sc}$, one may write,

$$< N(E_F) > = \frac{\int_0^{\Delta_{sc}} N(E) dE}{\int_0^{\Delta_{sc}} dE}$$

In the UD side, where the PG amplitude can be greater in magnitude compared to the SCG, we get (using the triangular PG model as described by Eqn. 1),

$$< N(E_F) > = \frac{\int_0^{\Delta_{sc}} \frac{N_0 E}{E_g} dE}{\int_0^{\Delta_{sc}} dE} = \frac{N_0 \Delta_{SC}}{2E_g} \qquad (3)$$

For the second condition, $E_g \leq \Delta_{sc}$, of the model gap, as shown in Fig. 1, we obtain,

$$< N(E_F) > = \frac{\int_0^{\Delta_{sc}} \frac{N_0 E}{E_g} dE}{\int_0^{\Delta_{sc}} dE} = N_0 \left(1 - \frac{E_g}{2\Delta_{SC}}\right) \qquad (4)$$

The characteristic pseudogap temperature, $T^*$ can be expressed as, $T^* \equiv E_g/k_B$ [11, 12, 17 – 19, 26]. Therefore, one can express the superpair density from Eqns. 2, 3, and 4 as follows,

$$\rho_s(p) = 2.29 N_0 k_B \frac{T_{c0}^2}{T^*(p)} \qquad \text{for } T^* > 2.14 T_{c0}$$

$$\qquad (5)$$

$$\rho_s(p) = N_0 k_B \left(2.14 T_{c0} - \frac{T^*(p)}{2}\right) \qquad \text{for } T^* \leq 2.14 T_{c0}$$

It follows from Eqns. 5 that within the proposed scenario, the hole content dependence of the superpair density arises from the *p*-dependent PG energy scale.

We have used Eqns. 5 to calculate the doping dependent superpair density of YBCO and Ca substituted YBCO. It should be noted that reliable independent estimate of $N_0$ does not exist in the literature. Therefore, we have calculated the normalized superpair density. A large body of experimental studies has demonstrated that $\rho_s(p)$ becomes maximum at $p \sim 0.19$ [15, 31, 39] where the PG vanishes quite abruptly. Hence we have fixed it to unity and calculated $\rho_s(p)$ with respect to this value at other hole concentrations. It is worth noting that the PG energy scale (and consequently $T^*$) does not depend on the level of Ca substitution in Y(Ca)BCO and depends solely on the number of doped holes in the $CuO_2$ planes [11, 12, 19]. It is also important to realize that $T^*(p)$ is insensitive to the crystalline state of the material [12, 19]; except for highly disordered compounds [20, 40], $T^*(p)$ is same in bulk single and polycrystals and thin films [12, 19] for a given value of *p*.

The $T^*(p)$ values for pure YBCO and Ca doped YBCO compounds are taken from prior published sources [11, 12, 19, 20]. Estimated values of normalized $\rho_s(p)$ obtained by employing Eqns. 5 are presented in Fig. 2.

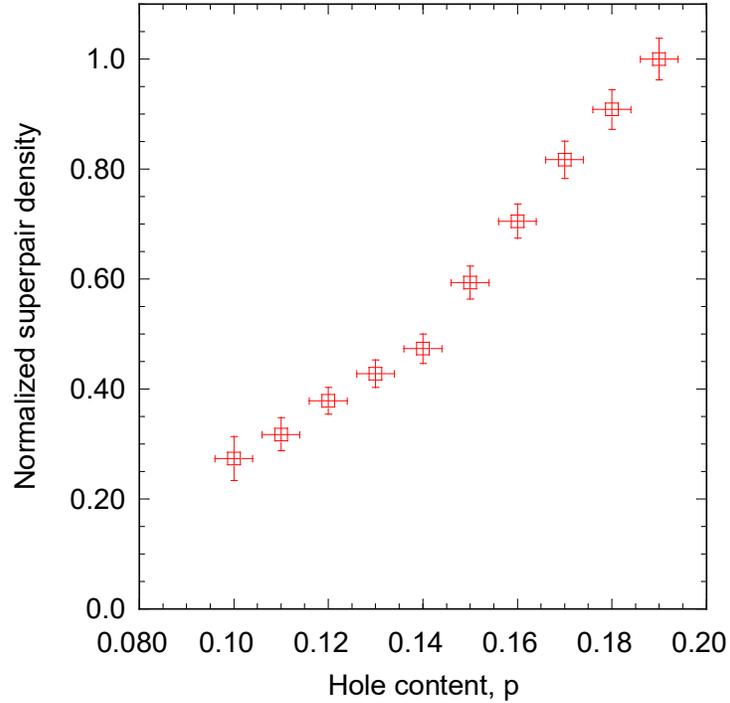

Figure 2: Variation of the normalized superpair density with number of doped hole content in the $CuO_2$ plane of YBCO. The hole contents are accurate within ± 0.004. The errors in the normalized superpair density come primarily from the uncertainty in the values of $T^*(p)$ [9, 11, 16, 19].

**3. Superpair density and the critical current density**

$J_c$ can be enhanced by increasing the vortex pinning force per unit volume. However, $J_c$ cannot be increased indefinitely even if it were possible to prevent vortex motion completely. There is an intrinsic limit to the maximum achievable supercurrent density. This limiting value is termed as the depairing current density, $J_{dp}$ [41]. It sets an upper limit to $J_c$. This depairing critical current density is an intrinsic characteristic of superconductors. It is directly related to the phase stiffness of the superconducting wavefunction. Actually the observed critical current density is primarily determined by this depairing current density in cuprates, particularly in YBCO and Ca substituted YBCO [9, 42].

The depairing current density, $J_{dp}$, corresponds to the current density at which the kinetic energy of the Cooper pair and the condensation energy become equal. $J_{dp}$ can be expressed as [43]

$$J_{dp} = \frac{\phi_0}{3\sqrt{3}\pi\mu_0\lambda^2\xi} \tag{6}$$

where $\phi_0$ is the flux quantum, $\mu_0$ is the permeability of vacuum and $\xi$ is the SC coherence length.

For high-$T_c$ cuprates $J_{dp}$ is approximately $10^9 \text{A/cm}^2$ [42]. Currently, the critical current density, $J_c$, reaches about $5-10\%$ of the Ginzburg–Landau depairing current density at 4.2 K for the best high-$T_c$ specimens [43]. Nevertheless, it has become evident from a variety of hole content dependent critical current density studies that it is this depairing contribution which sets the low-$T$ limit of $J_c$ [8 – 10, 42, 44].

The critical current density varies strongly with temperature. As temperature increases the extrinsic effects associated with defects of different nature within the compound start to dominate the temperature dependent $J_c$. The extrapolated zero-temperature $J_c$ at zero magnetic field, $J_0$, gives the true reflection of the intrinsic depairing contribution [9, 10]. In this investigation, we have presented the zero temperature critical current density of high-quality c-axis oriented thin films of Y(Ca)BCO as a function of hole content. The thicknesses of these epitaxial thin films lied within 2800 ± 300 Å. Hole content was varied for fixed level of Ca substitution via oxygen annealing under different temperatures and partial pressures. $J_0$ was extracted by fitting the hole content dependent zero-field critical current density employing the following relation [3, 45]

$$J_{c0}(t) = J_0(1-t)^n \tag{7}$$

where, $t = (T/T_c)$, is the reduced temperature and $J_{c0}$ is the zero-field critical current density obtained from the *M-H* hysteresis loops via the modified critical state formalism [9, 46]. Value of the exponent, $n$, in Eqn. 7 is dependent on the level of anisotropy, defect distribution, microstructure, and level of homogeneity in chemical composition [3, 9, 10]. Magnetic field was applied along the *c*-direction. Therefore, the critical current flowed in the *ab*-plane of the compounds. Details regarding the samples used in this study and magnetization measurements can be found in Refs. [9, 12, 47]. The extrapolated values of $J_0$ together with their normalized values are presented in Table 1 for a large number of Y(Ca)BCO thin films. The hole contents reported in this paper are determined from the room-temperature thermopower measurements [11, 19, 48, 49] and also via the application of the widely employed parabolic $T_c$-$p$ relation [50]. The reported values are accurate within $\pm 0.004$.

Table 1
Zero-field and zero-temperature critical current density of $Y_{1-x}Ca_xBa_2Cu_3O_{7-\delta}$ thin films.

| Compound | Hole content ($p$) | Critical current density, $J_0$ ($10^6$ A/cm$^2$) | Normalized critical current density |
|---|---|---|---|
| $YBa_2Cu_3O_{7-\delta}$ | 0.162 | 20.62 | 0.668 |
| | 0.146 | 14.84 | 0.481 |
| | 0.102 | 5.99 | 0.194 |
| $Y_{0.95}Ca_{0.05}Ba_2Cu_3O_{7-\delta}$ | 0.184 | 30.88 | 1.000 |
| | 0.170 | 26.04 | 0.843 |
| | 0.156 | 22.10 | 0.716 |
| | 0.123 | 12.10 | 0.392 |
| $Y_{0.90}Ca_{0.10}Ba_2Cu_3O_{7-\delta}$ | 0.198 | 23.73 | 0.831 |
| | 0.188 | 28.54 | 1.000 |
| | 0.162 | 23.50 | 0.823 |
| | 0.160 | 24.41 | 0.855 |
| | 0.126 | 13.08 | 0.458 |
| $Y_{0.80}Ca_{0.20}Ba_2Cu_3O_{7-\delta}$ | 0.201 | 17.08 | 0.921 |
| | 0.186 | 18.54 | 1.000 |
| | 0.166 | 17.01 | 0.917 |
| | 0.150 | 12.98 | 0.700 |
| | 0.144 | 12.03 | 0.649 |
| | 0.136 | 10.09 | 0.544 |

For meaningful comparison, we have plotted normalized superpair density and normalized zero-field, zero-temperature critical current density in Fig. 3 as function of number of doped holes in the CuO$_2$ planes. A clear correspondence between the calculated superpair density based on the triangular PG model and $J_0$ obtained from experimental critical current density is seen over an extended range of hole content from $p \sim 0.10$ to 0.19.

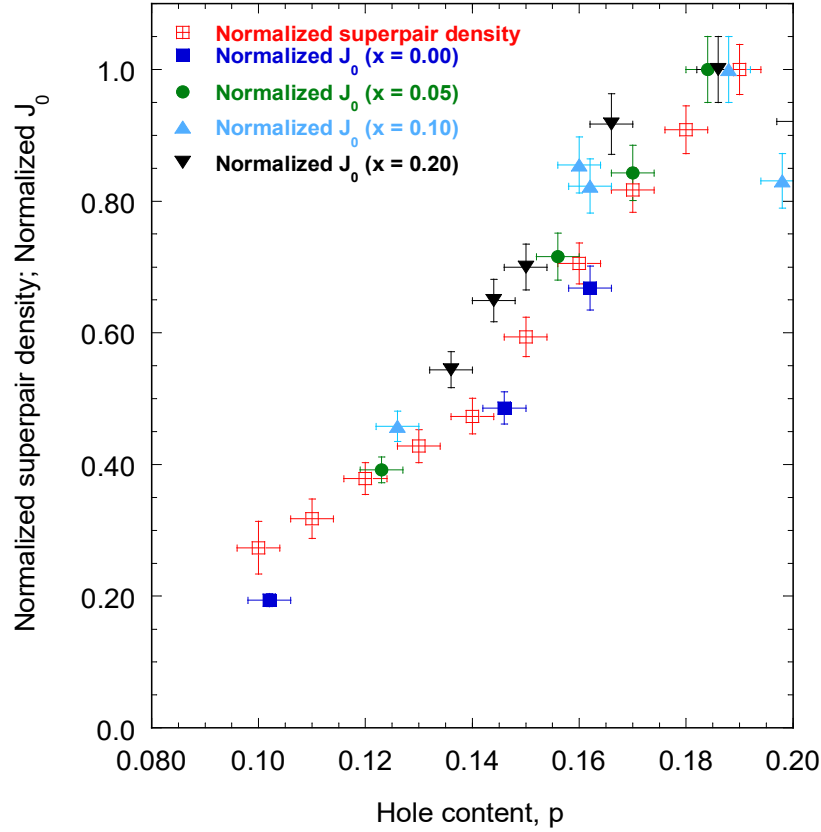

Figure 3: Variation of the normalized superpair density and normalized $J_0$ with hole content of Y(Ca)BCO superconductors.

## 3. Discussion and conclusions

In the preceding section, we have shown that a simple model based on a triangular PG in the QP spectral density pinned at the Fermi level can be used to estimate the superpair density quite easily. The normalized values of $\rho_s(p)$ closely follows the normalized $J_0(p)$ extracted from the experimental critical current density data over a substantial region of SC composition of Y(Ca)BCO. By definition, the superpair density should have the same physical significance as the superfluid density, $n_s$. As far as experimental studies are concerned, two of the most direct methods to obtain $n_s$ (actually $n_s/m^*$) are the magnetic penetration depth measurement and measurement of the muon spin resonance (μSR) depolarization rate [51, 52]. Both these measurements estimate $n_s/m^*$, where $m^*$ is the effective mass of the super-carriers. It is instructive to note that the both the μSR depolarization rate, $\sigma$ and $\lambda^{-2}$ are directly proportional to each other [15, 53], and the magnetic penetration depth is related to the superfluid density through $\lambda = [m^*/(\mu_0 n_s e^2)]^{1/2}$. Bernhard et al. [15] have calculated the hole content dependent superfluid density and SC condensation energy, $U_0$, of Y(Ca)BCO and $Tl_{0.5-y}Pb_{0.5+y}Sr_2Ca_{1-x}Y_xCu_2O_7$. We have plotted the results of $p$-dependent normalized $n_s/m^*$ of Y(Ca)BCO together with our results of normalized $\rho_s$ in Fig. 4. Considering the simplicity of the triangular PG model, the agreement between the theoretically estimated $\rho_s(p)$ and experimentally determined $n_s/m^*(p)$ is remarkable. The inset of Fig. 4 exhibits the hole content dependent normalized SC condensation energy of Y(Ca)BCO. The $p$-dependent features of the condensation energy is similar to that of $\rho_s(p)$ and $n_s/m^*(p)$. This has important consequence on the observed hole content dependence of the critical current density. The SC condensation energy shown below was calculated from the electronic heat capacity results for Y(Ca)BCO [22].

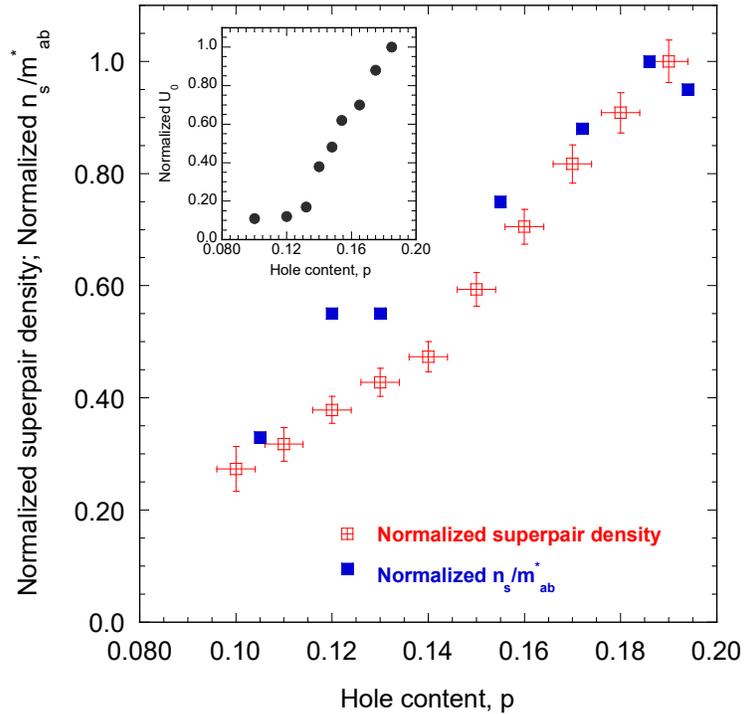

Figure 4: Main panel: Hole content dependent normalized superfluid density ($n_s/m^*_{ab}$) obtained from the μSR measurements [15] and the calculated normalized superpair density of Y(Ca)BCO. Inset: Hole content dependent normalized SC condensation energy, $U_0$, extracted from the heat capacity measurement [22] of Y(Ca)BCO superconductors.

The observed correspondence between the normalized $\rho_s(p)$ and the normalized $J_0(p)$ as illustrated in Fig. 3, can be understood from two different but related point of views. It follows from Eqn. 6 that the intrinsic depairing critical current density is directly proportional to $\lambda^{-2}$, and therefore, to $n_s/m^*_{ab}$, which in turn is a measure of the superpair density as shown in Fig. 4. It is worth noticing that $J_{dp}$ also varies with the SC coherence length $\xi$. But like the SC energy gap, the variation of $\xi$ with hole content is weaker than that of $\rho_s(p)$ or $n_s/m^*_{ab}(p)$. Therefore, the hole content dependent behavior of $J_c$ is dominated by the $p$-dependence of $\rho_s$ or $n_s/m^*_{ab}$.

It is known that the magnetic flux lines are pinned at locations (pinning sites) where the SC order parameter is partially or almost completely suppressed. Under this situation the pinning energy of the vortex core reveals itself as the energy barrier to the dissipative movement of the flux line and gives a measure of the flux activation energy $U_a$ [54]. It is this activation energy which sets the values of $J_c$ and the irreversibility magnetic field [45, 54] of a superconductor. Employing a simple heuristic scaling, Yeshurun and Malozemoff [55] and Tinkham [56] have found that $U_a \sim H_c^2$, where $H_c$ is the thermodynamical critical magnetic field. On the other hand, the SC condensation energy can be expressed as $U_0 \sim H_c^2$, which in consequence implies that, $U_a \sim U_0 \sim H_c^2$ [45, 54]. Furthermore, the SC condensation energy can also be expressed as $U_0 = \langle N(E_F) \rangle \Delta_{sc}^2$. This expression for $U_0$ establishes a direct link between vortex dynamics with a characteristic energy scale $U_a$, and the superpair density $\rho_s(p)$ (= $\langle N(E_F) \rangle \Delta_{sc}$). The arguments presented here are quite general in nature and do not depend significantly on the precise nature of the mechanism leading to Cooper pairing in a particular type of superconductor. For example, we have shown recently that the variation of the critical current density of heavy fermion superconductors (HFSCs) [57] with pressure follows strikingly similar pattern to the $p$-dependent variation of the $J_c$ of hole doped cuprates in Ref. [10]. The common thread is the pressure dependent variation in the SC condensation energy in the HFSCs and its $p$-dependent variation in the hole doped cuprates.

Within the proposed scenario, both $J_0(p)$ and $\rho_s(p)$ are maximized at a hole content ($p \sim 0.19$) where the PG vanishes, as found experimentally [11, 18, 19, 22, 31 – 35, 58]. For further overdoping, both these parameters decrease [8, 9, 15, 39, 42]. The origin of this reduction in the superfluid density in the deeply OD side is still unclear [39]. There is no PG in this particular side of the $T$-$p$ phase diagram. This behavior in the OD side implies that electronic correlations of different type (from the correlations giving rise to the PG in the UD to slightly OD region) competes with superconductivity and weakens the Cooper pairing correlations.

As far as the estimation of $\rho_s(p)$ based on the triangular PG model is concerned, there are a few points which needs some further elaboration. We have used a single, $p$-independent value of the SC energy gap given by $\Delta_{sc} = 2.14 k_B T_{c0}$ (with $T_{c0}$ = 93 K) for the calculations over a range of hole content from $p$ = 0.10 – 0.19. This approximation is reasonable for the hole contents $p$ > 0.12 [36, 37]. For Y(Ca)BCO compounds with lower level of hole content, the SC gap exceeds this value. This enhanced SC gap reduces the estimated value of the superpair density to some measure (at the level of ~ 5 – 10%). To account for this reduction, we have incorporated the appropriate error bars to the $\rho_s(p)$ of the two most underdoped compounds in Figs. 2 – 4. The triangular PG model assumes a flat EDOS outside the QP spectral energy gap region. This assumption is supported reasonably well by the $p$- and $T$-dependent coefficient of electronic heat capacity data [21, 22, 36].

It is instructive to point out that the $p$-dependent superfluid densities of Y(Ca)BCO, Tl$_{0.5-y}$Pb$_{0.5+y}$Sr$_2$Ca$_{1-x}$Y$_x$Cu$_2$O$_7$, Bi2212, and LSCO demonstrate almost identical features [15, 39], which imply that hole content dependent PG plays the prime role in determining this parameter in the hole doped high-$T_c$ cuprates and the model calculations presented in this study can readily be extended to other families of hole doped cuprate superconductors.

To summarize, we have employed a simple PG model to calculate the superpair density of high-$T_c$ cuprates. The estimated normalized superpair density shows clear correspondence to the experimental critical current density as a function of number of added holes in the CuO$_2$ planes of Y(Ca)BCO over a wide range of compositions. This strongly supports that the overall $p$-dependent behavior of the zero-temperature and zero-field critical current density is set by the depairing contribution. To check the efficacy of the model estimate of $\rho_s(p)$, we have compared it to the experimentally measured $n_s/m^*_{ab}(p)$ of Y(Ca)BCO. A very good agreement has been found.

**Data availability**

The data sets generated and/or analyzed in this study are available from the corresponding author on reasonable request.

**Author Contributions**

S. H. N. designed the project and wrote the manuscript. N. A. and S. H. N. performed the theoretical analysis. Both the authors reviewed the manuscript.

**Additional Information**

**Competing Interests**

The authors declare no competing interests.